\documentclass[lettersize,journal]{IEEEtran}
\usepackage{amsmath,amsfonts}
\usepackage{moresize}
\usepackage{algorithmic}
\usepackage{algorithm}
\usepackage{array}
\usepackage{booktabs}
\usepackage[caption=false,font=normalsize,labelfont=sf,textfont=sf]{subfig}
\usepackage{textcomp}
\usepackage{stfloats}
\usepackage{url}
\usepackage{verbatim}
\usepackage{color,soul}
\usepackage{graphicx}
\usepackage{cite}
\begin{document}

\title{A novel approach to differential expression analysis of co-occurrence networks for small-sampled microbiome data}

\author{Nandini Gadhia (University of Liverpool, STFC Hartree Centre) \\
Michalis Smyrnakis (STFC Hartree Centre) \\
Po-Yu Liu (Royal Veterinary College, School of Medicine, National Sun Yat-sen University, Taiwan) \\
Damer Blake (Royal Veterinary College) \\
Melanie Hay (Royal Veterinary College) \\
Anh Nguyen (University of Liverpool) \\
Dominic Richards (STFC Hartree Centre) \\
Dong Xia (Royal Veterinary College) \\
Ritesh Krishna (IBM Research)

\thanks{}
\thanks{\today}

\markboth{TCBB DRAFT PAPER \today }%
{Shell \MakeLowercase{\textit{et al.}}: A Sample Article Using IEEEtran.cls for IEEE Journals}}

\IEEEpubid{}

\maketitle

\begin{abstract}
Graph-based machine learning methods are useful tools in the identification and prediction of variation in genetic data. In particular, the comprehension of phenotypic effects at the cellular level is an accelerating research area in pharmacogenomics. Understanding the effect of drugs or disease on the underlying bio-network could facilitate future drug development and improvement of precision medicine. 

In this article, a novel graph theoretic approach is proposed to infer a co-occurrence network from 16S microbiome data. The approach is specialised to handle datasets containing a small number of samples. Small datasets exacerbate the significant challenges faced by biological data, which exhibit properties such as sparsity, compositionality, and complexity of interactions. Methodologies are also proposed to enrich and statistically filter the inferred networks. The utility of the proposed method lies in that it extracts an informative network from small sampled data that is not only feature-rich, but also biologically meaningful and statistically significant. Although specialised for small data sets, which are abundant, it can be generally applied to any small-sampled dataset, and can also be extended to integrate multi-omics data.

The proposed methodology is tested on a data set of chickens vaccinated against and challenged by the protozoan parasite \textit{Eimeria tenella}. The raw genetic reads are processed, and networks inferred to describe the ecosystems of the chicken intestines under three different stages of disease progression. Analysis of the expression of network features derive biologically intuitive conclusions from purely statistical methods. For example, there is a clear evolution in the distribution of node features in line with the progression of the disease. The distributions also reveal clusters of species interacting mutualistically and parasitically, as expected. Moreover, a specific sub-network is found to persist through all experimental conditions, representative of a ‘persistent microbiome’. A clustering algorithm is implemented on the network to demonstrate its utility for downstream analysis. 

\end{abstract}

\begin{IEEEkeywords}
bioinformatics, microbiome, 16S, network, graph, workflow, omics, Eimeria. co-occurences, bootstrapping
\end{IEEEkeywords}

\section{Introduction}

\IEEEPARstart{T}{he} rapid acceleration in the development of both high throughput sequencing (HTS) technologies and hardware capabilities at the start of the millennium decreased genome sequencing costs \cite{b31} and opened opportunities to investigate genomes and populations at the molecular level. Analysing the variation of ecological communities and their interrelations is naturally well-suited to graph-based machine learning methods. This is because they have interpretable clusters, are dynamic in response to disturbances (disease or treatment) and offer explainable visualisations \cite{b33}. 

This research area lies in the intersection of two disciplines, machine learning and biology; as such, standard network construction and analysis tools are not specialised to handle bioinformatic data. The data possesses distinctive challenges such as small sample size, compositionality and sparsity \cite{b33}. On the other hand, standard phylogenetic diversity techniques  such as characterising the alpha and beta diversities \cite{b24} in a sample, are not powerful enough to reveal nuances in the interplay of species interactions \cite{b26}. In recent years, studies have successfully applied graph-based approaches to genomic sequencing data \cite{b12, b26, b29}, albeit utilising large samples of sequencing data. This work draws inspiration from these approaches, but suggests a novel approach to infer, filter and analyse networks, designed specifically for small data sets. Experiments that produce small data sets (data sets with significantly fewer samples than inferred features) are common in biology, due to the ethical and financial challenges associated with genome and deep amplicon sequencing and processing. 
 
Here, a bespoke method is proposed to tackle these challenges. The method proposes that co-occurrence networks are constructed to represent the microbes in an ecosystem and their interrelations. A relation (edge) between two species or sequence types (nodes) in a network is defined by co-presence in a sample. To describe the strength of the relationships, not all of which are equal, the edges are weighted by a function of the relative abundance of the microbes they share a connection with. The role of the species as components of the system are expressed by introducing node and edge attributes with both biological and statistical relevance. Unlike traditional approaches, which apply filtering steps before network construction, this method recommends Monte-Carlo methods to generate a population from which the significance of the observed population can be determined. By exploring the statistical properties of the simulated population, spurious artifacts in the network representation of the data are identified and discarded. This methodology statistically enhances the quality of the network embedding.

The proposed approach is applied to 16S rRNA gene sequences from the hypervariable V3-V4 regions of 88 chickens sampled in an anticoccidial vaccination \cite{b2} trial.  

As a prerequisite to network construction, bioinformatic preprocessing transforms reads into tabular data; a table detailing the abundance of each Amplicon Sequence Variant (ASV) in each sample. This preprocessing comprises of a series of specialist steps, usually using single-purpose community-built tools, stitched together to form a pipeline \cite{b34}. A bioinformatic workflow automates this pipeline, and also ensures reproducibility, scalability, hyper-parameter tracking,  version control and reference annotation. Such a bioinformatic pipeline was developed as part of this work. It is written in a workflow manager called Nextflow \cite{b5}, which is extensible to other omics data types, and is high performance computing (HPC) enabled. As well as enabling portability by handling software installation by containerisation, the workflow pipeline prevents effort duplication by the researcher applying each tool separately. 

The rest of this article is structured as follows. The next section discusses related work and current approaches to network inference. The issues encountered with small-sampled biological data sets are highlighted. The proposed solution to network inference and filtering, is detailed in Section \ref{section:construction}. Section \ref{section:data} introduces the 16S data used to demonstrate the application of the proposed methodology, before Section \ref{section:processing} which details the pre-processing workflow developed to process this data from raw sequenced reads. Section \ref{section:application} analyses the filtered networks, which includes the application of a clustering algorithm on the inferred and filtered networks. Finally, Section \ref{section:conclusions} summarises this work and discusses future work. 


\section{Current approaches}
\label{section:network_considerations}

 It is clear that as bioinformatics grows from a data sparse to data rich field of study, analysis needs to extend beyond the alpha and beta diversity analyses \cite{b24} that, until recently, represented most sequence-based microbial community analysis. Although biological data suffers from compositionality and sparsity, which challenges network traversal algorithms, it also exhibits various topological properties that make it suitable for exploration using network techniques \cite{b23}.

Ecological networks are prone to cluster formation \cite{b25}, the detection of which is well researched with network techniques \cite{b54}.  The detection of such subnetworks can provide valuable insight into disease etiology or therapeutic responses \cite{b53}.

Bionetworks are sensitive to perturbation by infection or treatment \cite{b52}, wadvocating the analysis of changes in networks as a means to capture responses of the underlying microbiota. This could facilitate target identification and drug discovery \cite{b55}. Microbial species are highly connected, and these connections are representative of their interactions \cite{b52}. Thus network methods can be used to investigate the interactome for the identification and prioritization of disease-causing candidate genes \cite{b56}. Nonetheless, the characteristics of bioinformatic data must be taken into account when applying traditional graph techniques. Specifically, this work focuses on network construction and inference. A discussion of considerations for the analysis of genetic data follows.

\subsection{Filtering}
Studies by Lam et al.\cite{b57} and Ngedwa et al. \cite{b58} are exemplary of studies which filter data with lower prevalences before analysis. The motivation for these filtering methods, such as rarefaction, arises from the requirement to identify species that appear as a result of imprecision in the experimental tools. Despite bioinformatic pipelines implementing contaminant removal methods, spurious associations can still remain due to sequencing errors and instrument uncertainties \cite{b50}. Prevalence filtering is partly successful in addressing this ubiquitous issue since it is likely that the species above the imposed threshold is at least impoverished in spurious associations. However, while prevalence filtering is easy to implement, thresholds imposed are largely arbitrary and conventions vary from lab-to-lab, thus lessening the reproducibility and rigour of an analysis.

Nearing et al. highlight that different methods can produce significantly different results \cite{b42}. Consequently, a filtering method that employs statistical methods for robustness would be preferred. This is especially true for small-sampled data since there are not enough samples to justify the classification of species as rare based on prevalence alone. Rusell et al.\cite{b43} successfully apply prevalence analysis to gut microbiome data, but the number of samples is in surplus of 400. This large number of samples allowed for more confidence in this method since over 500 species remain after 5\% prevalence filtering. For smaller sampled data, such a threshold is more likely to lead to significant loss of information.

\subsection{Sparsity, Compositionality and Conditional Independence} Typically, network inference defines an interrelation between two features from the abundances measured in multiple statistically independent observations. Treating species or sequence types as independent inherently ignores the compositional nature of genetic data; many standard analysis techniques have been shown to be invalid when applied on compositional data \cite{b49}.  Genetic data are compositional because counts measured by high throughput sequencing are not absolute counts. Instead their magnitudes are limited by arbitrary sequencing depths imposed by thresholds on the measurement instruments. Gloor et al. \cite{b18} warns of the necessity to account for compositionality at every stage of the analysis. 

Microbiome data are inherently sparse \cite{b45}. Such zero inflation arises for two reasons. Firstly, biological/structural zeros arise when a species is truly absent in an ecosystem. These zeros are further supplemented by sampling zeros - those that arise as a result of low sequencing depth and sampling variation. \cite{b35}. Both effects couple to give extremely sparse initial matrices.

Many current methods use linear regression to ascertain relationships between representative species and environmental factors \cite{b48}. However, linearly transformed data are only appropriate under the assumptions of normality and constant variance, which are violated by zero-inflated data \cite{b46}. Similarly, the use of logistic regression \cite{b47} leads to a loss of information since zeros are treated as non-events. However, species that vanish in specific conditions (biological zeros) could be important biomarkers in understanding the effects of an infection. 

New methods, such as SparCC \cite{b26}, have been proposed to account for sparsity and compositionality in large-scale networks. This is accomplished by implementing a log-ratio transformation for every pair of correlated taxa.  While the correlation based method is useful for detecting indirect associations, it ignores the concept of conditional independence. Furthermore, for smaller data, the SparCC algorithm is likely to produce a matrix that is not positive definite (the data used by the authors of the method to exhibit the application of SparCC contains 200 samples \cite{b26}). Such a matrix cannot be partially inverted to account for conditional independence.

Another method that addresses both compositionality and conditional independence, SPIEC-EASI \cite{b27} uses the centered log-ratio transformation on relative abundances. It uses both neighbourhood selection \cite{b36} and graph lasso  \cite{b26}. However, it does not converge for small biological datasets since it assumes the data are distributed as a uni-variate or a multivariate Gaussian. This assumption is problematic in this instance because the data are proportional, extremely right skewed and over-dispersed since there are not enough samples to apply the central limit theorem \cite{b41}.

The small number of samples is especially a problem because the number of inferred features much larger than the number of samples as a result of the high diversity in 16S data. Even tools which assume a log-normal distribution over-emphasise the weight of zeros, since they are not designed in a bioinformatics context. An over-emphasis on zeros implicates an absence of species (biological zeros), when the presence of a zero might simply be a result of the limitations on high throughput experimental techniques(sampling zeros) \cite{b33}. Small sample size is a common feature of biological datasets since the practical processes required to create or extract data are often expensive, involved and pose ethical challenges. Therefore it is imperative to develop a method that accounts for these characteristics for inherently sparse matrices, whilst accounting for compositionality.

\subsection{Higher order interactions} Genetic networks are diverse and complex. Although the presence of an edge between two nodes/species suggests a dependence between the species, their relation might in fact be dependent on a cross-feeding relationship with a third species \cite{b51}. These higher order interactions are often neglected in network analyses with constant edge weights. Graph traversal techniques such as subgraph learning and motif detection allow for the identification of higher order relationships.

\section{Network construction}
\label{section:construction}
A graph-network is a collection of objects (called nodes or vertices) that are connected by edges. Mathematically, a graph is  represented as $\mathcal{G}= (\textbf{V}, \textbf{E})$
where $\textbf{V}$ denotes the set of vertices in $\mathcal{G}$ and $\textbf{E}$ denotes the set of edges of $\mathcal{G}$. The number of nodes is $N =\|\textbf{V}\|$, unless otherwise specified. For our dataset, the vertices(nodes) represent the features inferred by the preprocessing phase. Specifically, a feature is an ASV (Amplicon Sequence Variant), which is a collection of 16S rRNA sequences that have a certain percentage of sequence divergence.

An adjacency matrix $\textbf{A}$ of a graph is a square matrix that defines the connections present in a graph. The component $A_{ij}$ is 1 if the vertices $v_i$ and $v_j$ share a connection, and 0 if not.

 Here, using an approach initially similar to \cite{b12}, a microbial co-occurrence network is constructed for each experimental condition. The adjacency matrices for each sample are aggregated and normalised such that an edge between species m and n in the network signifies that species m and n are both present in at least one of the samples. Since each network represents a unique microbial landscape for each experimental condition, characterising the changing co-occurrence and co-exclusion patterns serves as an important step to understanding infection-linked or vaccine-linked imbalances. The number of samples in which species m and n occurred together is recorded. This ensures information is not lost as a result of aggregation of the samples. 

\subsection{Edge attributes}
Co-occurrences are useful for capturing the existence of an interaction between species but are agnostic to the nature or strength of the interaction. In bioinformatic data, species can have both direct and indirect interactions, both of which are important to capture \cite{b29}. Therefore, edge weights are introduced to capture how similarly connected species respond to abiotic effects.

To this end, a weight is given to each edge as follows: 
For species m and n, with relative abundances $r_m$ and $r_n$. Let $x_{mn}$ be the ratio of their relative abundances. 
\begin{equation}
    x_{mn} = \frac{r_m}{r_n}
\end{equation}

Then define weight of the edge between them:
\begin{equation}
    w_{mn} = \left(\frac{x_{mn} + \frac{1}{x_{mn}}}{2}\right)^{-1}
\end{equation}

Weighting the edges in this way further informs the embedded relationships between the species, since it captures the similarity in the relative abundances of the two connected species. While relative abundances alone capture the magnitude of the presence of the species, the edge weights additionally emphasise species which might be low in abundance, but change in congruence with each other, and might therefore indicative of the molecular machinery at play. An edge weight close to 1 suggests the ratio of the abundances is close to 1, thus capturing mutualism between species \cite{b28}, while a lower edge weight captures commensalism or perhaps parasitism.

\subsection{Node attributes}
The existence of the species (represented as nodes) and their connections is supplemented with information about the connectivity and importance of the species as part of the whole ecological network. These node attributes enable us to further inform the graph to capture a system-level perspective. This enables us to understand how connected a species is and how significant these connections are. The attributes can also be used to improve the explainability of algorithms. 

Attributes derived from node properties include  average relative abundance, average degree, average weighted degree, betweenness centrality, average number of co-occurrences. For each node biological attributes are also recorded in the form of the taxonomic labels inferred in the last step of the preprocessing pipeline. This enables the investigation of node properties across various experimental conditions. 

\subsection{Filtering}
Issues with arbitrarily filtering data were discussed in Section 2. These methods are particularly concerning because they would fail to preserve rare species. These unique species might be indicative of the microbiome architecture of a particular condition. Furthermore, the interpretation of presence–absence investigations on over-filtered data would raise concern, for instance if used to define specific sub-networks such as a core microbiome \cite{b59} (a sub-network of species integral to ecosystem function). 

Hence we propose a method to detect spurious associations and false positives via statistical methods. For each experimental condition, there exists an underlying probability distribution that describes the variation of species presence and abundance in that condition. Since this intrinsic distribution cannot be accessed explicitly, Monte-Carlo methods test whether an observation of species fits this distribution. To this end, testing the observed edges allows for the identification of spurious species that are likely to be a product of inaccuracies in the experimental or preprocessing method. 

Let $\Tilde{\mathcal{G}} $ be the original inferred co-occurrence network for a given experimental condition. Furthermore, let $\Tilde{w}$ be the observed value of an attribute. For ease of notation, $\Tilde{w}$ represents the following cases. 
\begin{itemize}
    \item For node level analysis, \[
\Tilde{w_i} = \frac{\sum_{j \in N(i)} w_{ij}}{k_i}
\] is the observed average weight of the edges connected to a node \(i\).

Here \(N(i)\) represents the set of neighbors of node \(i\), and \(k_i\) is the degree of node \(i\), i.e., the number of edges connected to node \(i\).

    \item For edge level analysis, $\Tilde{w_{ij}}$ is the observed weight of the edge between species i and its neighbour species j.
\end{itemize}

Synthetic data are created for each experimental condition. Re-sampling from the original data ensures that each bootstrapped network is derived from the same underlying probability distribution as the observed data, despite this distribution being impossible to ascertain explicitly. 

Therefore, for each experimental condition, n bootstrapped graphs $\mathbf{G_B} = [\mathcal{G}_1, \mathcal{G}_2... \mathcal{G}_n]$ are created. These synthetic graphs are created and populated with attributes using the same methodology as described in section \ref{section:construction}, but from synthetic ASV tables. The synthetic ASV tables are generated by randomly selecting 10 samples without replacement from the 10 observed samples in each experimental condition. For the central limit theorem to apply, n ought to be sufficiently large. 

Then confidence intervals can be defined, and empirical p-values calculated to test that the null hypothesis that the average of this same attribute in the bootstrapped population is the observed value. 

\begin{center}
$\boldsymbol{H_0} :   \mu_{w,bootstrapped} = \Tilde{w}$ \\
$\boldsymbol{H_1} :   \mu_{w,bootstrapped} = \Tilde{w}$
 
\end{center}

\textbf{Node level significance testing}: Here the statistical significance of individual species is tested. According to the central limit theorem \cite{b41}, the distribution of the mean weight of a species is normally distributed. A confidence interval of two standard deviations captures 96\% of the population. Species whose observed average weight falls outside the two-sigma range could be considered spurious and pruned from the network. 

\textbf{Edge level significance testing}: A representation of the interactions of an ecostystem needs to carefully account for the nature of microbes; species are known to compete for similar resources and rely on cross-feeding for survival \cite{b44}. Hence, spurious associations could be inferred by associating a response in a particular microbe with a collective response. Nodes left isolated by edge-level pruning can be discarded or left as isolated nodes.

The method describes as above is applied to each edge in the network and a confidence interval defined. Edges outside the 2$\sigma$ confidence interval are discarded.

\subsection{Revealing a persistent microbiome}
\label{section:persistent_microbiome}

With confidence in the relevance of the proposed co-occurrence networks, an intersection of species that persists through all the conditions can be identified. Analysing the differential expression of this sub-network is of great interest since it is an ecosystem of species which is present in all species, but with varying node and edge attributes. 

In the literature, a `core microbiome' is defined as the members common to two or more microbial assemblages associated with a habitat \cite{b19}. Identifying the core species (or core ASVs) is important because the commonly occurring organisms are likely to be critical to the function of that type of habitat, and therefore identifying changes in this core network could be indicative of the changes to immune response.

\section{Experimental data}
\label{section:data}

16S rRNA gene sequencing data was collected by The Royal Veterinary College from the hypervariable V3-V4 regions of 88 Cobb500 broiler chickens (NCBI BioProject accession number PRJNA990995). The chicken were infected with \textit{Eimeria tenella} and treated during an anticoccidial vaccination experiment  trial \cite{b2}. These data exhibit the characteristics of biological data discussed above. The data classified in groups by their source (caecal lumen content / caecal tissue) infection statuses (unchallenged / challenged), vaccination statuses (unvaccinated / mock vaccinated / vaccinated) and the number of days post-infection (DPI) the sample was taken at. The categorisation of the samples are detailed in Table \ref{figure:data_table}.

\begin{table}[ht]
\caption{A table detailing the data set used. For the analysis in article, the experimental conditions marked with an asterix, are used.}
\label{figure:data_table}
\centering
\begin{tabular}{@{}p{1cm}p{2.5cm}p{1.5cm}p{1cm}p{0.5cm}@{}}
\toprule
\textbf{Caecal source} & \textbf{Vaccination} & \textbf{Infection} & \textbf{Samples} & \textbf{DPI} \\
\midrule
Content & Unvaccinated    & Challenged   & 10 & 6  \\
Content & Unvaccinated    & Unchallenged & 10 & 6  \\ 
Content & Mock-vaccinated & Challenged   & 10 & 6  \\ 
Content & Vaccinated      & Challenged   & 10 & 6  \\ 
* Tissue  & Unvaccinated    & Challenged   & 10 & 6  \\ 
* Tissue  & Unvaccinated    & Unchallenged & 10 & 6  \\ 
Tissue  & Mock-vaccinated & Challenged   & 10 & 6  \\ 
Tissue  & Vaccinated      & Challenged   & 10 & 6  \\ 
* Tissue  & Unvaccinated    & Challenged   & 8  & 10 \\
\bottomrule
\end{tabular}
\end{table}

The challenge infection was initiated by an oral dose of 15,000 \textit{Eimeria tenella} (Houghton strain) oocysts administered 21 days after hatch). The parasite was chosen because it can cause coccidiosis, a pervasive disease which causes annual losses exceeding USD twelve billion  to the farming industry by way of loss of productivity and high treatment costs \cite{b1}. Consequently, the vaccine trial used an experimental yeast-based anticoccidiosial vaccine \cite{b2} in some of the individuals. Most of the chickens were sampled six DPI, but one group was challenged and sampled ten DPI. Thus, the network behaviour is investigated over a quasi time series.

Understanding the behaviour of the \textit{Eimeria} parasite and its impact on enteric microbiomes at the cellular level is crucial for the control and prevention of coccidiosis, and potentially also the development of novel treatments. While the method used in this case study is generalisable to any small sampled 16S rRNA data, the real world application of investigating the \textit{Eimeria tenella} infection on the host enteric microbiota further motivates this study. The data consisted of 176 paired-end FASTQ files for this analysis. The next section describes the processing pipeline developed, which reads the sequencing data, agglomerates it, infers taxa and performs initial visualisation.

\section{Pre-processing}
\label{section:processing}
The recent acceleration in data driven microbiome research has resulted in  a lack of scalable and reusable infrastructure for processing such data. Although a community consensus has settled on a set of state of the art tools \cite{b8} for processing an omics data set, the effort to use each of these tools separately is duplicated by each researcher. One of the common challenges in complex bioinformatics workflows is data provenance capture\cite{b37} i.e. storing the details of each workflow execution including the software versions, dependencies and parameters. This is important to ensure transparency into the underlying processes and accountability in the case of inconsistent results \cite{b35}. It is also important that analyses are reproducible so that they can be bench-marked with current best practices at any time \cite{b37}.

For this reason, the tools required for this study and its dependencies are containerised in both Docker V.20.10.21 \cite{b39} and Singularity V.1.0.2 \cite{b38}. A reactive workflow framework (Nextflow V.22.10.6 \cite{b5}) stitches together these containers, and is optimised to leverage the high performance computing facilities at the Hartree centre. Nextflow \cite{b5} uses a data driven approach and comes with its own domain scripting language (DSL2), which provides an abstraction layer between the logic of the pipeline and the execution. It is particularly useful that the architecture of Nextflow allows for the full data set to be segregated such that each sample can be processed separately, and then combined after preprocessing. This naturally achieves considerable speedup for the most computationally expensive tasks (sequence variant inference). As a result, it can be executed by any user on multiple platforms. For this study, the workflow is supported for execution on a local machine or on the Hartree Centre's supercomputing system, Scafell Pike, which is a Bull-Atos Sequana X1000 supercomputer. The pipeline is optimised and parallelised such that execution is fast for small data sets on most machines.

\subsection{Quality control} The paired reads are passed through a quality control tool, FASTQC V.0.11.9 \cite{b3}, and aggregated using the MultiQC extension. FastQC detects potential deviations from expected frequencies and summarizes the distribution of quality scores \cite{b4}.

\subsection{Trimming} From the quality scores, the low quality sections of the read can be truncated using Cutadapt using the plugin within QIIME2(V.2022.8) \cite{b6}. Trimming results in shorter reads but with a higher average per base quality score. It also ensures all reads are the same length. For our data, a truncation length is specified at the 210th base pair for the forward read and the 240th base pair for the reverse read.

\subsection{Infer representative sequence variants} QIIME2(V.2022.8) \cite{b7} converts the tabular data from the raw rRNA reads. The q2-demux plugin demultiplexes samples to remove redundancies (samples with more than two expected errors). DADA2 \cite{b8} infers samples from amplicon data with single-nucleotide resolution, which reduces the likelihood for stochastic variation to be interpreted as biological variation. Pooling is used to improve the detection of variants that are infrequent in individual samples but numerous across all samples. Finally, chimeric sequences are removed and forward and reverse reads merged into a single paired read.

\subsection{Assigning taxonomies} The representative sequences of the ASVs are assigned to taxa via the q2-feature-classifier plug in. This performs supervised classification \cite{b13} using an alignment-based taxonomy consensus method based on VSEARCH. \cite{b10}. The reference database used was the SILVA 138 (2019) database  \cite{b9}, which is comprehensive, quality checked and regularly updated. The resultant taxonomy table represents the bacterial diversity of the microbes in the sample, assigning a kingdom, phylum, class, order, family, genus and species to each sample (where possible).

This pipeline outputs 728 distinct ASVs across the 88 samples with corresponding representative sequences and assigned taxa. 

\section{Application to dataset}
\label{section:application}
The 176 FASTQ files are imported into the workflow described in section \ref{section:processing}. The networks are constructed for each condition group (Table (\ref{figure:data_table}) as described in Section \ref{section:construction}; the results follow.

 This article will investigate chickens sampled at three stages of infection - unchallenged chickens, challenged chickens sampled at six DPI and challenged chickens sampled at ten DPI. All groups are unvaccinated.

The first of these groups is also sampled at 6 DPI, but because it is unchallenged it acts as the zeroth point in the quasi-time series. The last of these conditions represents chicken in which the disease is almost resolved, and hence allows us to investigate lasting effects of the disease on the composition of the ecosystem. 

\subsection{Filtering}

The bootstrap population (n=1000) is generated as previously described in the Methodology section. 

\begin{figure}[ht]
\centering
\includegraphics[scale=0.2]{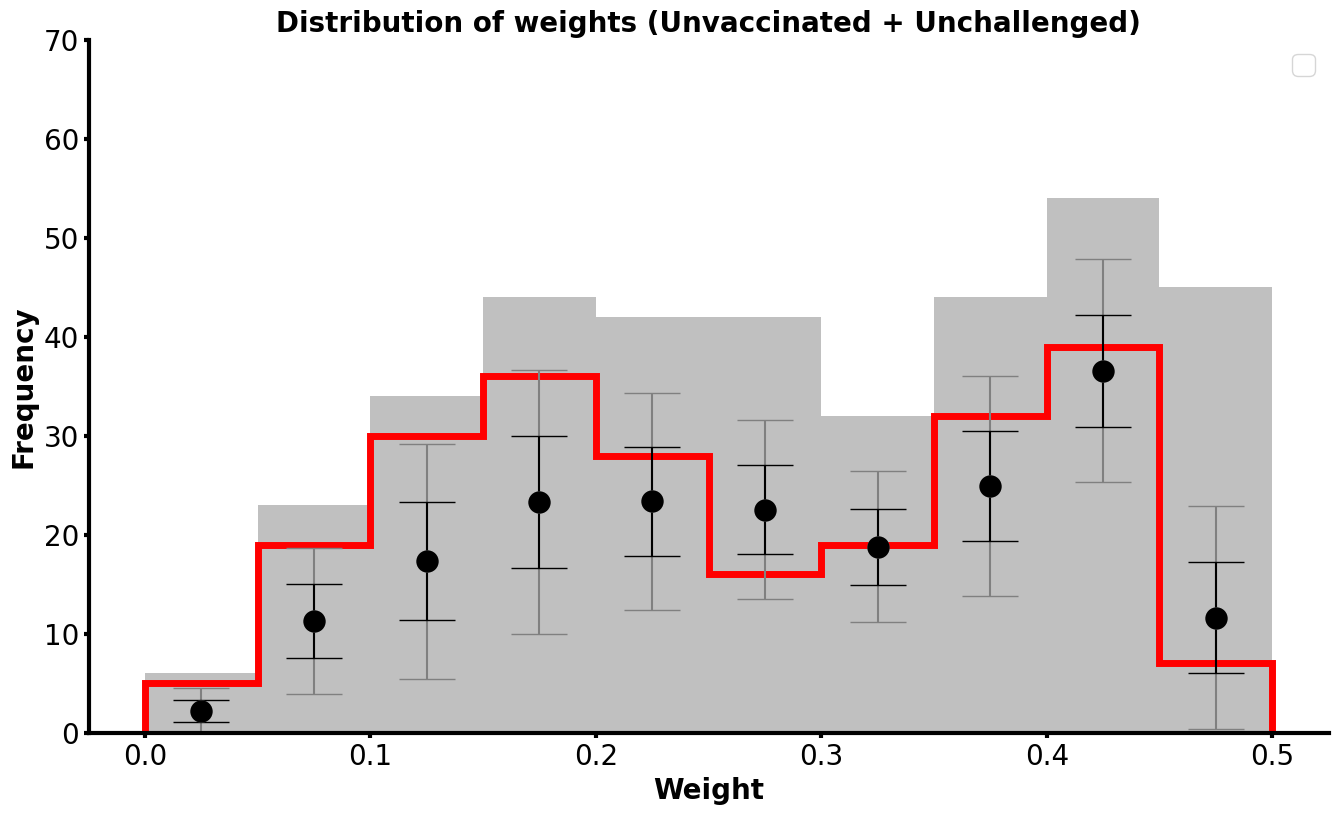}
\caption{Bootstrapped weight distribution against actual data distribution of weights. The mean and standard deviations (one and two) from the mean are plotted as error bars. Data from unvaccinated and unchallenged condition.}.
\label{figure:bootstrap_weights_5}
\end{figure}

The distribution of weights for the synthetic data and observed data are plotted in Figure \ref{figure:bootstrap_weights_5} for the three time series conditions. It provides an indication into the utility of this method for pruning spurious species. Clearly, a number of species are beyond the range of two standard deviations of the mean frequency in the group.

\subsubsection{Node-level filtering}
To investigate the contributions of individual species further, hypothesis testing can be performed on the species individually. The species with an observed average weight outside the constructed confidence interval and all their connected edges would be discarded from the network. 

Table \ref{figure:species_filt_table} details the number of species that would be filtered for the data set used in this work.

\begin{table}
\centering
\caption{A table comparing the number of species filtered under different methods}

\begin{center}
\resizebox{\columnwidth}{!}{%
\begin{tabular}{|c|c|c|c|c|}
\hline
\textbf{Condition} &\textbf{Original number of nodes}& \textbf{Abundance filtered (1\%)}& \textbf{Statistically filtered (1$\sigma$)}& \textbf{Statistically filtered (2$\sigma$)} \\

\midrule
Unchallenged & 231 & 142 & 106 & 0 \\
Challenged (6 DPI) & 218 & 142 & 70 & 0 \\
Challenged (10 DPI) & 134 & 72 & 26 & 0 \\
\midrule

\end{tabular}
}
\label{figure:species_filt_table}
\end{center}
\end{table}

\subsubsection{Edge-level filtering}
For this work, filtering is done by calculating p-values for the observed network compared with the bootstrap population. A two-tailed hypothesis test for each data group is performed at the 5\% significance level. Equivalently, confidence intervals can be constructed for each edge present in a network; edges with an observed weight outside the confidence interval are pruned from the network. 

\begin{table}
\centering
\caption{Results from edge-level filtering}

\begin{center}
\resizebox{\columnwidth}{!}{%
\begin{tabular}{|c|c|c|c|c|c|c|}
\hline
\textbf{Condition} &\textbf{Original number of edges}& \textbf{Filtered number of edges} & \textbf{SPIEC-EASI edges} & \textbf{Original number of nodes}& \textbf{Filtered number of nodes} & \textbf{SPIEC-EASI nodes} \\

\midrule
All samples & 80152 &  2758 & 4802 & 740 & 265 & 740 \\
Unchallenged & 15715 & 795 & - & 231 & 93 & - \\
Challenged (6 DPI) & 17536 & 132 & - & 218 & 58 & - \\
Challenged (10 DPI) & 6064 & 42 & - & 134 & 52 & - \\
\midrule

\end{tabular}
}
\label{figure:edge_filt_table}
\end{center}
\end{table}

\begin{figure}[ht]
\centering
\includegraphics[scale=0.35]{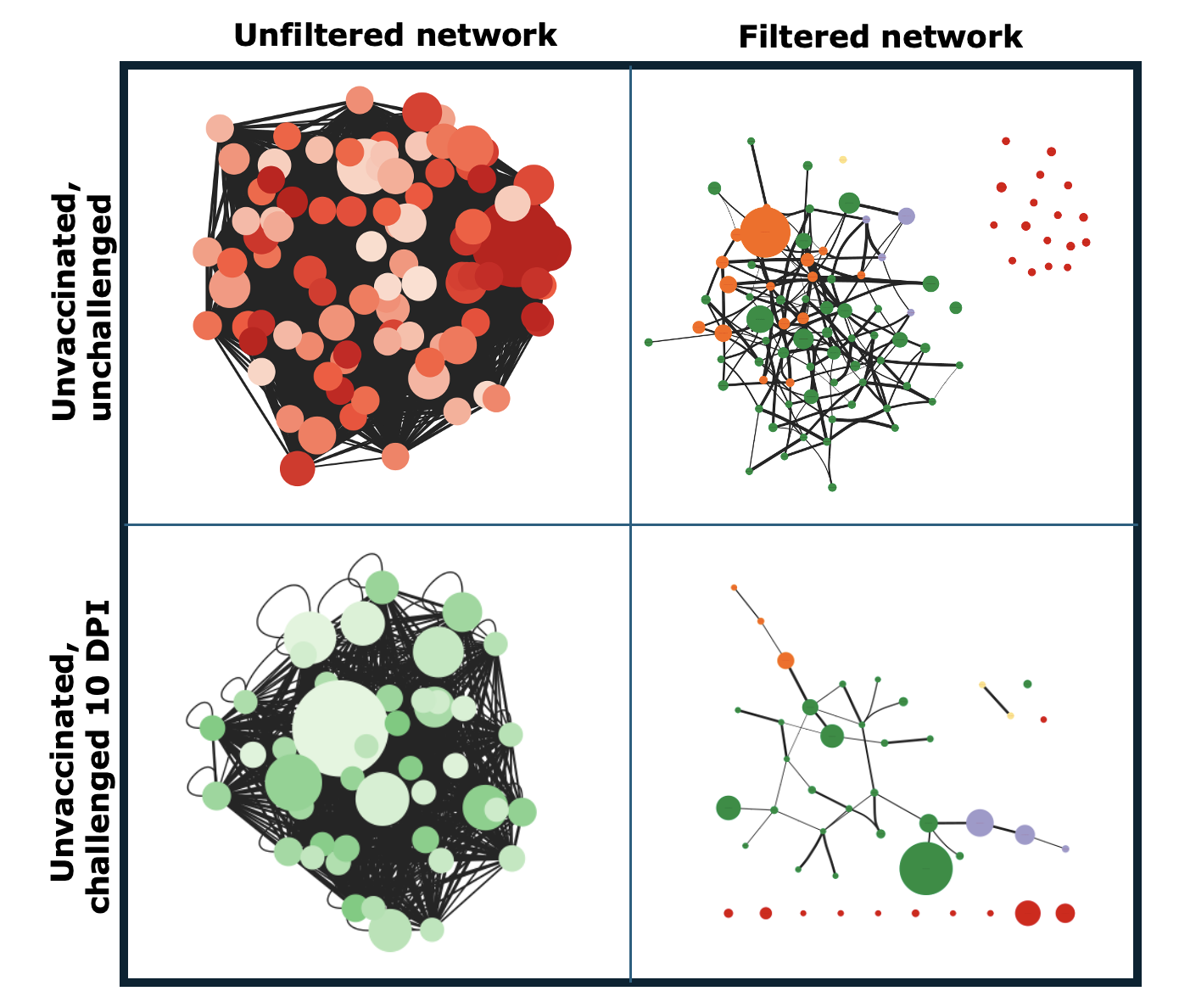}
\caption{A visualisation of the co-occurence network of samples in two conditions before and after the application of the proposed pruning methodology. Before filtering, community clustering algorithms \cite{Su_2010} are unable to group connected components into distinct clusters. This hairball can be logically simplified using a variety of computational approaches which can help biologists understand a complex network in a manageable and biologically meaningful manner}.
\label{figure:edge_trim_network}
\end{figure}

Figure \ref{figure:edge_trim_network} displays the co-occurence networks of one sample in the unvaccinated, unchallenged cohort, and one sample in the unvaccinated, challenged 10 DPI cohort; both before and after application of the proposed pruning methodology. The networks are visualised using Cytoscape \cite{Shannon_2003} and the default Affinity Propagation clustering algorithm is implemented in all cases. Affinity propagation \cite{Frey_2007} is a clustering methodology that uses similarity computation and message passing between data points to identify cluster centres and assign data points to these clusters. It is hence particularly useful when the number or shape of clusters is not known in advance.

One way to investigate co-occurence networks is via clustering; the decomposition of interation networks into communities of densely interacting nodes. While the biological validity of the identified clusters requires further investigation, clustering algorithms identify functional modules and can identify important bio markers. Before filtering, the algorithm is unable to effectively apply clustering. Each node is placed into self contained, individual clusters. This is because spurious associations present in the graph mask discerning features within the graph. After filtering, four clusters are identified, suggesting that the removal of spurious associations allows for the emergence of more distinct communities within the networks.

\subsection{Comparison with other methods}

\subsubsection{Prevalence filtering}
\begin{figure}[ht]
\begin{center}
\includegraphics[width=0.6\linewidth]{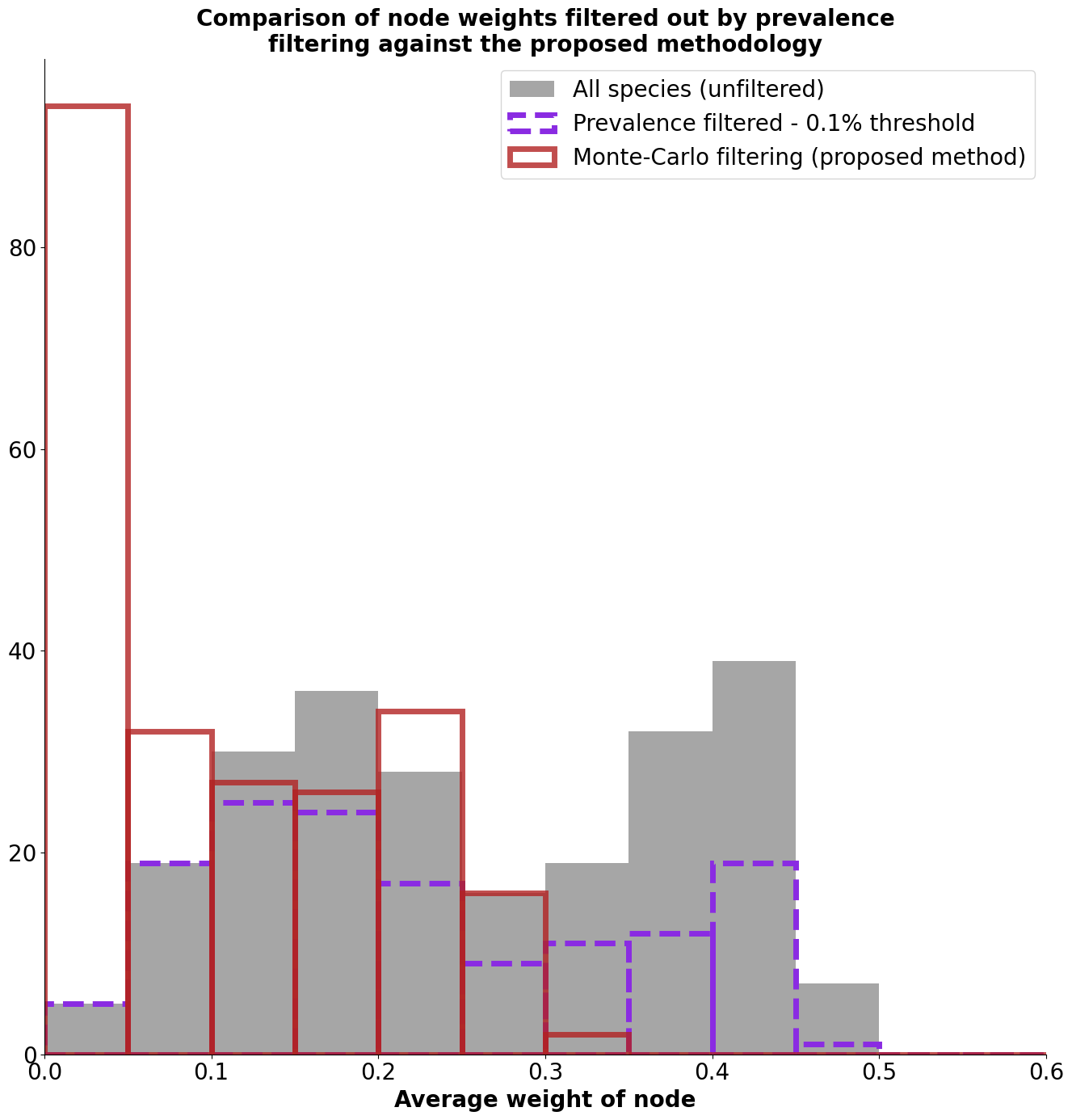} 
\end{center}
\caption{The distribution of the discarded weights in a typical filtering workflow filtering by abundance (0.1\% threshold) compared with the proposed methodology of filtering using statistical methods.}.
\label{figure:discarded_weights}
\end{figure}

Figure \ref{figure:discarded_weights} shows the weights of the species that would have been discarded in a typical prevalence filtering workflow. The plots use species from the unvaccinated and unchallenged condition. A 0.1\% threshold was placed on the relative abundances of the species `filtered'. This is a relatively conservative choice as many workflows choose to use a threshold as high as 5\% \cite{b8}. 

This further validates the necessity for a more conscientious to pruning for spurious associations. Clearly prevalence filtering discards species at larger weights and centralities that are likely to be crucial to explaining the interactions in an ecosystem. 

Furthermore, the proposed statistical approach has the advantage that the pruning process can be automated to filter species outside two standard deviations from the bootstrapped mean. Prevalence filtering methods require the threshold to be manually imposed based on the experiment's designs and goals. 

\subsubsection{SPIEC-EASI}

SPIEC-EASI, a network inference algorithm published in 2015, uses sparse inverse covariance estimation for infer ecological networks. It aims to take advantage of the proportionality invariance of relative abundance data and address the difficulties previously mentioned in wide data sets. SPIEC-EASI implements the following steps for network inference:

\begin{itemize}
    \item The ASV count data is taken as input. It is preprocessed and centered log-ratio (CLR) transformed. This addresses compositionality.
    \item An option of two methods is implemented for model inference neighbourhood selection using the MB method \cite{Meinshausen_2006} or inverse covariance selection using graph lasso \cite{glasso}.
    \item StARS \cite{liu2010stability} corrects for sparseness.
\end{itemize}

For our dataset, SPIEC-EASI fails to converge with either inference method for each separate experimental condition (10 samples). For the entire dataset, the MB method converges to a graph with 740 nodes and 4802 edges; 2663 of the 4802 edges have weights between -0.05 and 0.05. By construction, the weights in SPIEC-EASI are not directly comparable with the weights in the proposed methodology. Nevertheless, with over 55\% of edge weights not statistically differentiable from 0, these spurious interactions are over-represented in the SPIEC-EASI network, with the assumption that the weight between the nodes is meaningful in conveying the interaction between species. 

\subsection{Results}
\label{section:results}

\subsubsection{Analysis of networks}
\begin{figure}[ht]
\centering
\includegraphics[width=0.7\linewidth]{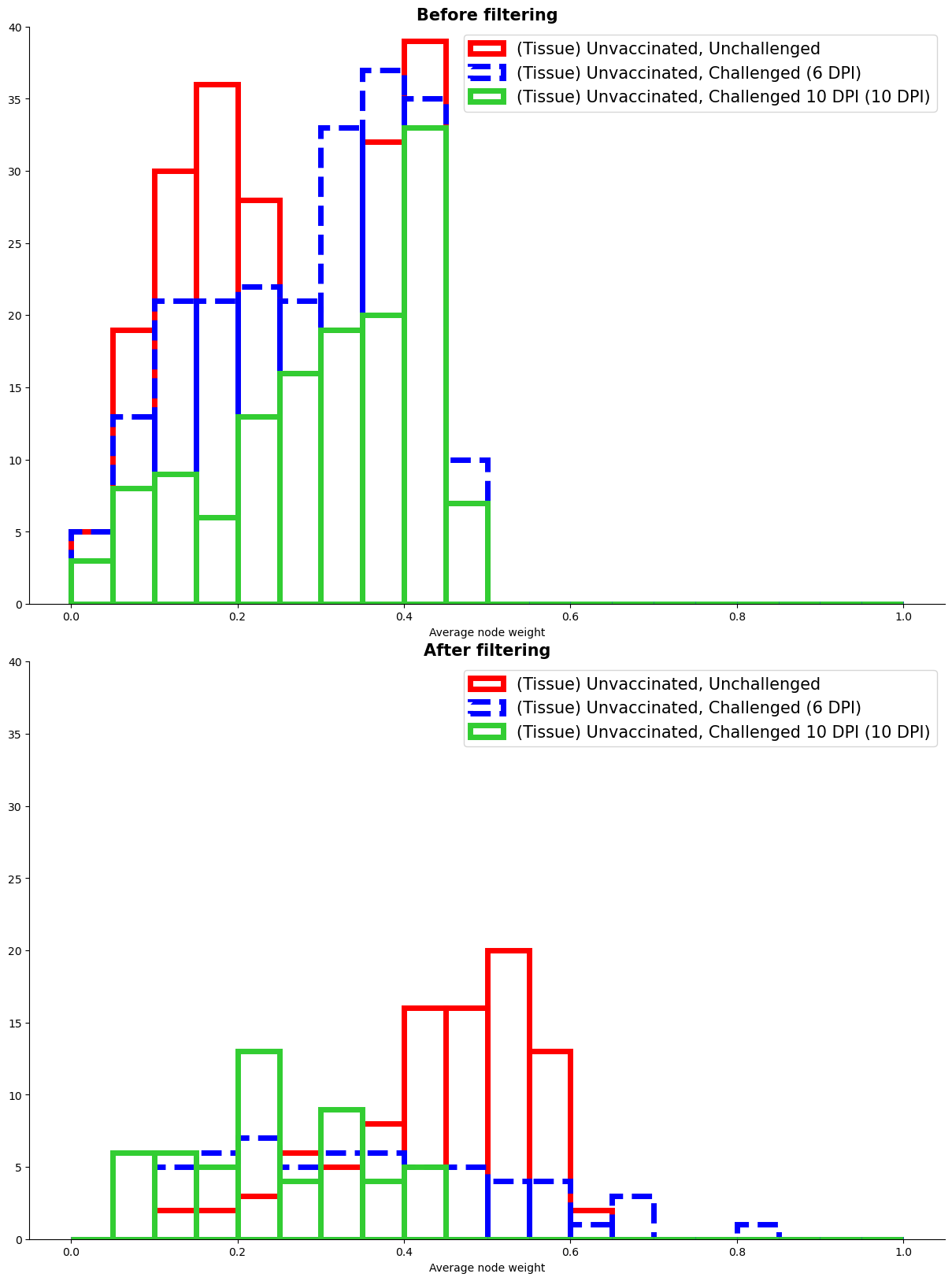}
\caption{Histograms showing average weight distribution for the unvaccinated species in three different stages of infection before filtering (top) and after filtering(bottom).}
\label{figure:hist_458_comb}
\end{figure}

Various properties of the networks can be plotted to extract insights from the entangled networks.  Figure \ref{figure:hist_458_comb} plots the average weight surrounded each node in a network, for the three experimental conditions of interest. This offers an insight into the ecosystem at node level, which translates biologically to a species level perspective.

The distributions before filtering appear bimodal, particularly at the earlier 2 stages of infection. From the two peaks, one at a lower weight($\sim$0.15) and the other at a higher weight ($\sim$0.4) . From this the species can be approximately classified into two groups; a group with weights of very different magnitudes to its neighbours, and a group with weights of more similar magnitudes. This supports our hypothesis that analysis of the distribution of weights would identify interactions of parasitism (the lower weight) and mutualism (the higher weight). 

To compare the distributions, we perform a Kolmogorov-Smirnov (K-S) test to determine whether the distributions are the same. The KS test statistic is defined as the maximum value of the difference between distribution A and distribution B’s cumulative distribution functions (CDF).

\begin{equation}
    \text{K-S test statistic} = \text{sup} \lvert F_A(x) - F_B(x) \rvert
\end{equation}

With the null hypothesis that the distributions are the same, the K-S test returns a p-value of 0.135 for the comparison of the unchallenged and challenged (6 DPI) conditions, and a p-value of $6.65 * 10 ^{-5}$ for the comparison of the challenged (6 DPI) and challenged (10 DPI) conditions. Therefore at the 5\% level of significance we can conclude that the unchallenged and challenged (6DPI) distributions come from the same underlying distribution, but between 6 and 10 DPI, the expression of the network changes significantly. 

Biologically, this might be interpreted as the progression of the disease escalating more rapidly between 6 and 10 days than between 0 and 6 days. It could also be perceived as the immune response kicking in and hence significantly impacting the ecosystem of the patient.

The bottom set of distributions in figure \ref{figure:hist_458_comb} shows the distributions of average node weight after filtering, for the three experimental conditions of interest. The K-S test results in a p-value of 0.005 for the comparison of the unchallenged and challenged (6 DPI) conditions, and a p-value of $8.89 * 10 ^{-4}$ for the comparison of the challenged (6 DPI) and challenged (10 DPI) conditions. At the 5\% significance level, we reject the null hypothesis for both tests and conclude that all the samples come from different distributions. 

The differential expression of the distributions is particularly interesting for the sample taken at ten DPI i.e. when the infection was almost resolved. Phenotypically the resolution of the disease suggests recovery to full health, but the significantly left-skewed distribution of the edge weights compared to six DPI suggests ecological variation. Identifying the species responsible for this change and experimental verification of the same could lead to the identification of target species for novel vaccine development or improved resilience. 

The distributions can also be compared before and after filtering. As a natural consequence of filtering, many of the edges with lower edge weights are considered spurious, and hence discarded. This skews all the average weights towards the right, and all the distributions show greater spread. Particularly, the control group (unvaccinated and unchallenged) is skewed towards nodes having higher average weights, which indicates their connection with similar nodes. As the disease progresses, we see the distributions being skewed towards lower weights. 

\subsubsection{Persistent microbiome}

Distinguishing a persistent microbiome, as defined in section \ref{section:persistent_microbiome}, reveals 63 species for the unfiltered data across all conditions. For the conditions of interest in the quasi time series, the filtered data for the quasi time series contains a persistent microbiome of just 8 species. The taxonomic decomposition of these 8 species is detailed in \ref{figure:pesistent_taxa}.  

\begin{figure}[ht]
\begin{center}
\includegraphics[width=0.8\linewidth]{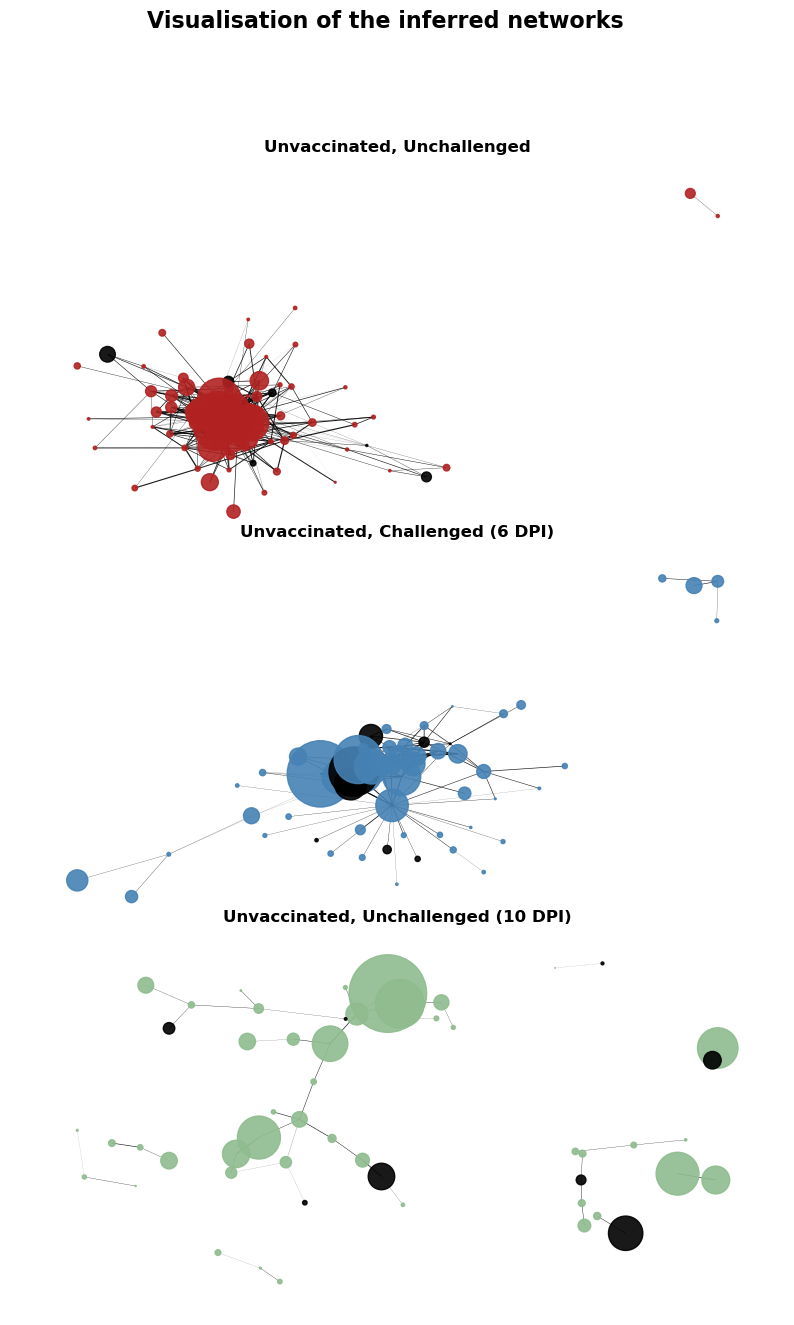}
\end{center}
\caption{Visualisation of the inferred networks. The persistent microbiome is distinguished in black, and their taxonomic decomposition detailed in Table \ref{figure:pesistent_taxa}}.
\label{figure:network_comb}
\end{figure}

The networks are visualised in Figures \ref{figure:network_comb}. The scale of the magnitude and the thickness is comparable between the three categories since their corresponding values are always normalised in the range (0,1]. The persistent microbiome is emphasised in black; despite the species being present in all conditions, there are evident differences in the abundances of the species (node size) and magnitude of edge weights (edge thickness). 

\begin{table*}
\centering
\caption{Taxonomic decompostion of the persistent microbome}

\begin{center}
\resizebox{\textwidth}{!}{%
\begin{tabular}{|c|c|c|c|c|c|c|c|}
\hline
\textbf{Node ID} &\textbf{Kingdom}& \textbf{Phylum}& \textbf{Class}& \textbf{Order} & \textbf{Family} & \textbf{Genus} & \textbf{Species}\\

\midrule
24 & Bacteria & Proteobacteria & Gammaproteobacteria & Burkholderiales & Burkholderiaceae & Burkholderia-Caballeronia-Paraburkholderia & - \\
31 & Bacteria & Proteobacteria & Gammaproteobacteria & Enterobacterales & Enterobacteriaceae & Escherichia-Shigella & - \\
32 & Bacteria & Patescibacteria & Parcubacteria & Parcubacteria & Parcubacteria & Parcubacteria & Uncultured parcubacteria \\
67 & Bacteria & Bacteroidota & Bacteroidia & Bacteroidales & Prevotellaceae & Prevotella & - \\
85 & Bacteroidota & Bacteroidia & Bacteroidales & Bacteroidaceae & Bacteroides & - \\
104 & Bacteria & Firmicutes & Clostridia & Oscillospirales & Oscillospiraceae & UCG-005 & - \\
105 & Bacteria & Firmicutes & Clostridia & - & -  & - \\
131 & Bacteria & Firmicutes & Clostridia & Oscillospirales & Oscillospiraceae & Oscillibacter & - \\

\midrule

\end{tabular}
}
\label{figure:pesistent_taxa}
\end{center}
\end{table*}

In contrast to the node-level perspective that the distributions in Figure \ref{figure:hist_458_comb} offer, these plots of the inferred networks offer a network level perspective by  illustrating the variation in edge weights and relative abundances for the ecosystems of each condition. For example, a significant increase in the relative abundance of a species in a resolved chicken compared to before and during infection could indicate the effects of immune response. On the other hand, depletion of other species could indicate a weakened system as a result of the disease. Since none of the chicken in these samples are vaccinated, these species could be candidates for further investigation for targeted vaccine development or for the development of a probiotic. 

From this systems-level perspective, there is a clear separation of the species as the disease progresses, with the network in the 'challenged 10 DPI' condition significantly more disjointed than the network in the 'unchallenged' condition. This change in the network topology indicates biological properties that might be expected from an ecosystem stressed by a parasite. This supports related work \cite{eimeria} that finds enteric dysbiosis in the bacteria of \textit{Eimeria}-infected animals.

\subsubsection{Clustering}
\label{section:clustering}

A hierarchical clustering algorithm was implemented on a feature table of node attributes for each experimental condition. The attributes considered were average abundance, degree centrality minimum and maximum weight, range weight and median weight. The average co-occurences and occurences were deliberately dropped from the feature matrix because their distributions are bipartite, and hence would dominate the clustering algorithm without revealing biological intuition. The dissimilarity matrix is calculated using the Bray-Curtis index \cite{bray}, since this is a metric often used in biology and ecology since it accounts for compositionality. 

The number of clusters, k, is determined by plotting the gap statistic \cite{gap_statistic} for various numbers of clusters. The gap statistic compares the total intra-cluster variation for different values of k with their expected values under null reference distribution of the data. For the unchallenged condition, the gap statistic peaks at 5 clusters. We therefore choose k=5 for all the conditions for ease of comparison. 

\begin{figure}[ht]
\begin{center}
\includegraphics[width=0.5\linewidth]{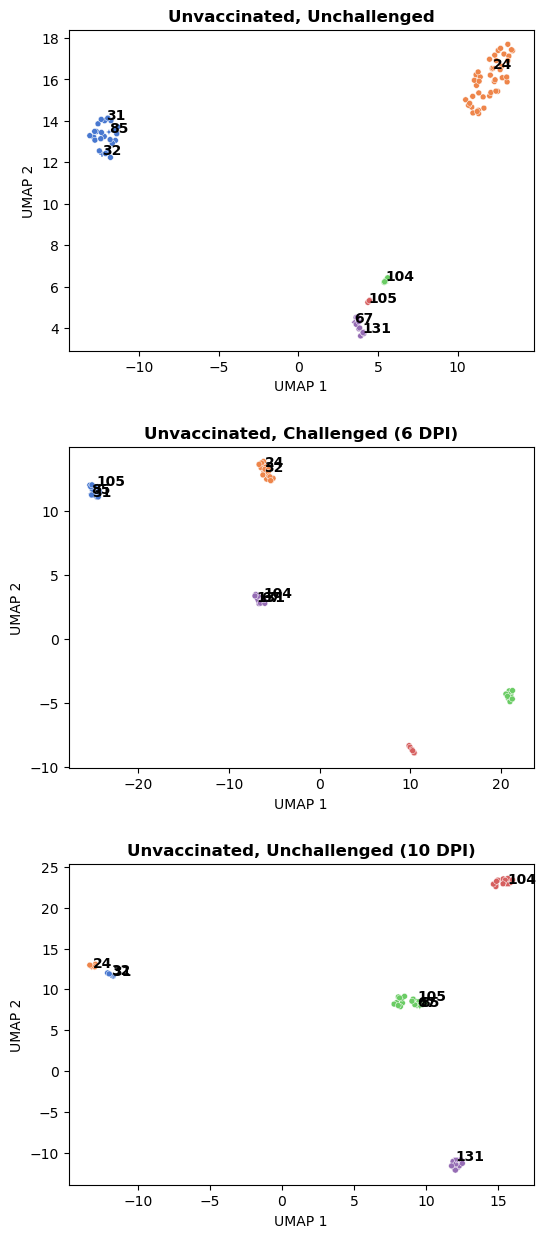}
\end{center}
\caption{Clustering(k=5) using hierarchical clustering with Bray-Curtis distance for unvaccinated and challenged (10 DPI) condition. 2D projection for visualisation using UMAP, with each cluster distinguished by colour. The species in the persistent microbiome are numbered. Their taxa are detailed in Table \ref{figure:pesistent_taxa}.}
\label{figure:clustering_comb}
\end{figure}

Figure \ref{figure:clustering_comb} visualises the clustering, projected into 2 dimensions using a manifold projection technique, UMAP \cite{mcinnes2020umap}. The numbered species identify the species in the persistent microbiome. 

Species 31 and 104 are clustered together in the first two stages of disease progression, but are separated at 10 days post infection. Similarly, species 131 and 104 are clustered together in the diseased states but not in the unchallenged condition. This could be an indicator of a distressed ecosystem. Species 67 and 104 are clustered together in the first and last stages of disease, but not in between, at 6 days post infection. If the disease is thought to be nearly resolved at 10 days post infection, this connection could be indicative of a healthy ecosystem. Furthermore, species 24 and 32 are grouped together in all stages except the last one, though their separation could be an artefact of 5 being a sub-optimal number of clusters for that experimental condition. More robust biological analysis of the connection between these species could be the topic of further investigation. 

\begin{figure}[ht]
\includegraphics[width=\linewidth]{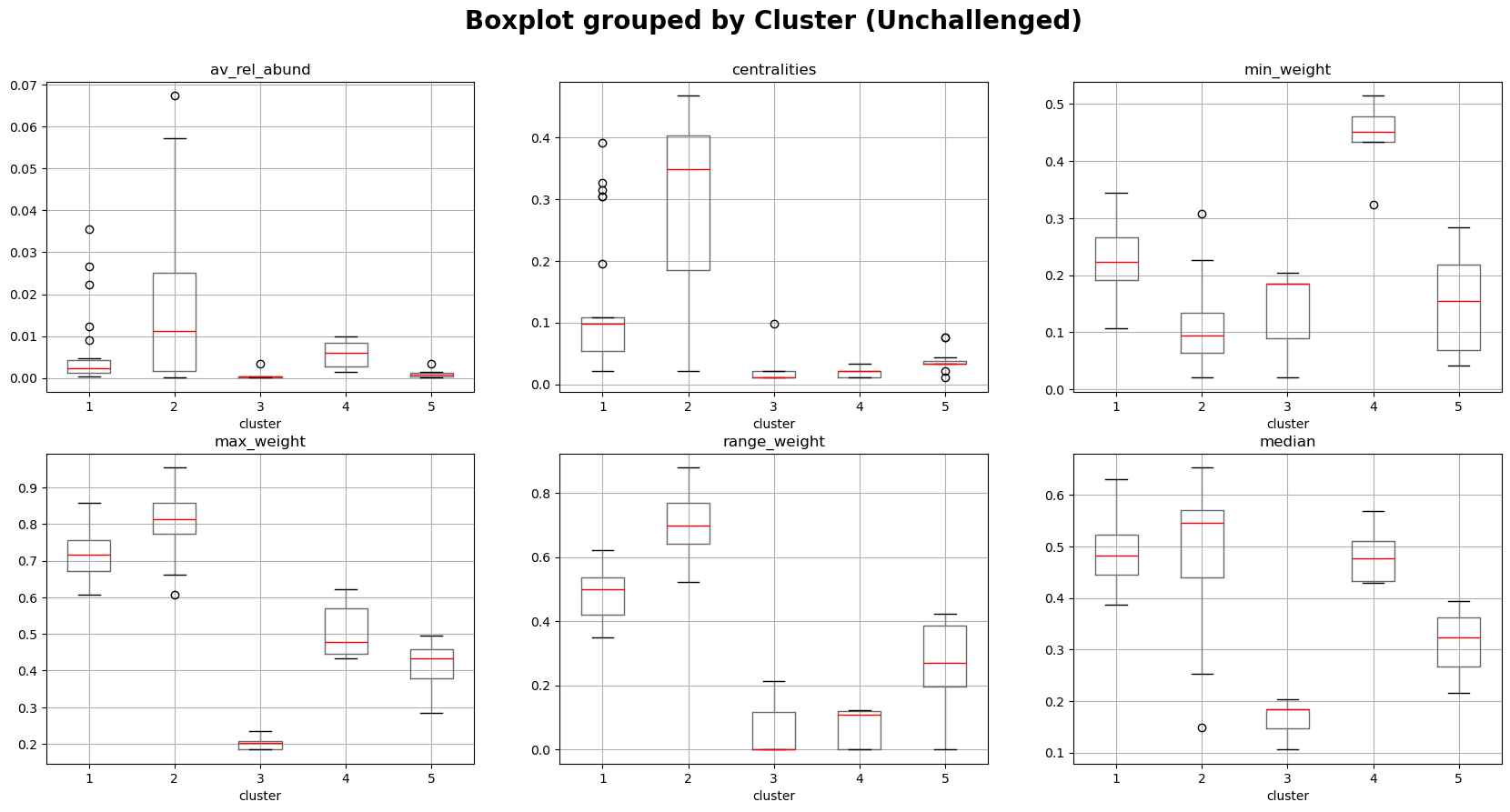}
\caption{Distribution of node attributes for each cluster identified in the unvaccinated and unchallenged condition.}.
\label{figure:clustering_boxplot_UU}
\end{figure}

The box plots in figures \ref{figure:clustering_boxplot_UU} plot the distributions of the features for each cluster within the unchallenged condition. The range and median weights are particularly helpful in discriminating between clusters. On the other hand, the average relative abundances between clusters is fairly similar for all experimental conditions. The distributions of the relative abundances and the centralities of the 5 clusters are consistent across experimental conditions.

The methodology proposed in this work is specialised for small-sampled data sets, where state of the art tools struggle to converge to a biologically meaningful network representation. While it is generalisable to any 16S data set, it is untested on larger networks.

The results of this analysis exhibit clear indications that the differential expression of the networks derive biological insights from data driven methods, notably the extraction of the core microbiome. These distributions are general systems perspectives and offer limited insights into the movement of individual species across the stages of disease progression.

Future work aims to integrate biological intuition to investigate individual species dynamics more robustly. Identifying particular nodes which contribute significantly to the changing expression of the enteric microbiome and could offer interesting biological comprehension with applications in precision medicine. The clusters identified in section \ref{section:clustering} need to be validated against biological enterotypes. The inferred network embeddings will be enhanced with functional pathway information derived from metagenomic inference. This additional data might enhance the utility of the clusters detected by machine learning methods.

The architectures developed during this work are scalable and lend to the integration of multi-omics data, such as transcriptome and metabolome. The bio-networks will be extended to multi-layered graphs to the incorporate additional data, illustrate latent relationships across data sets and allow for more holistic analysis. There is also scope for further statistical methods to be applied such as random-walk traversal and the creation of probability distributions to analyse the changing interactions across various experimental conditions.

\section{Conclusion}
\label{section:conclusions}
Sequencing data often consist of a small number of samples, and usually exhibit challenges such as compositionality and sparsity and spurious observations. Therefore, traditional network and statistical methods must be applied with caution and with biological relevance in mind. To tackle many of these challenges, a methodology to infer and enhance a bio-network from 16S rRNA gene sequences is proposed. The proposed solutions can be used for filtering spurious species and interactions and can be combined or used separately for any 16S microbiome data set. Co-occurrence matrices for each experimental condition were constructed, and the inferred networks enhanced with node and edge attributes. The proposed filtering methods were also demonstrated on the data at various scales: species (node) level and interaction (edge) level. 

The proposed method is tested on amplicon data from \textit{E.tenella} infected broiler chickens. The data from samples at three stages of disease progression showed significant variability as the  disease progressed and was resolved. Furthermore, a persistent microbiome was revealed across all experimental conditions. A hierarchical clustering algorithm implemented on the experimental conditions begins to reveal the dynamics of the species, in particular the persistent microbiome as the infection progresses. More robust biological analysis of the network dynamics could lead to interesting insights into the impact of the parasite on chicken ecosystems.

\section*{Acknowledgments}
This work was supported by the Hartree National Centre for Digital Innovation, a collaboration between IBM and STFC. The experimental data was produced with support from the Biotechnology and Biological Sciences Research Council (BBSRC, grant reference BB/P003931/1) and the Houghton Trust.

Special thanks to Matthew Marshall and Mark Birmingham at STFC Hartree Centre (Research Software Engineering) for invaluable contributions to the bioinformatics processing pipeline and software architecture.

This study was conducted in strict accordance with the Animals (Scientific Procedures) Act 1986, an Act of Parliament of the United Kingdom. All procedures were approved by the Animal Welfare Ethical Review Body (AWERB) of the Royal Veterinary College and the UK Home Office under the Project Licence PDAAD5C9D.

\section{References Section}
\bibliographystyle{IEEEtran}
\bibliography{sample-base}

\section{Biography Section}
\textbf{Nandini Gadhia}
Nandini Gadhia completed her B.Sc. (Hons) in Maths and Physics from the University of Warwick in 2020, followed by a MPhys. at University College London. She is currently a researcher at STFC Hartree Centre and a doctoral candidate at the University of Liverpool, focusing on ML for omics data.
\\

\textbf{Michalis Smyrnakis}
Michalis Smyrnakis obtained his B.Sc. from Athens university of Economics and Business  on Statistics, followed by an M.Sc. by research in Aston University on Pattern Analysis and Neural Networks. He obtained his PhD from Bristol University on game-theoretic learning. He is currently leading the Artificial Intelligence Group of Hartree National Centre. His research interest include Game Theory, Deep Reinforcement Learning, Graph Theory and Graph Neural Networks,  Deep Learning and Bayesian Inference. 
\\

\textbf{Po-Yu Liu}
Po-Yu Liu completed his B.Sc. (2012) in Biology and Public Health at Kaohsiung Medical University, followed by an M.Sc. (2014) in Zoology at National Taiwan University. He received his Ph.D. (2019) from Academia Sinica and National Taiwan University in systems biology, particularly the gut microbiota of folivorous flying squirrels. He is currently Assistant Professor at National Sun Yat-sen University with research focuses on understanding the gut microbiome forming mechanisms, K-mer-based algorithm development, and clinical microbiota research on fatty liver disease.
\\

\textbf{Damer Blake}
Damer Blake received his B.Sc. from Wye College, University of London, followed an M.Sc. and Ph.D from the University of Aberdeen. He is currently Professor of Parasite Genetics at the Royal Veterinary College with research focuses on genetics of parasites, especially apicomplexans, as well as enteric microbiomes, gut health, and pathogens of poultry. He is co-Director of the Centre for Emerging, Endemic and Exotic Diseases (CEEED) and Editor-in-Chief of the journal Avian Pathology.
\\

\textbf{Melanie Hay}
Melanie Hay received a BA from Rhodes University, and a BSc (Med) (Hons) in Medical Biochemistry and an MSc (Med) in Exercise Physiology from the University of Cape Town. She received her PhD from Aberystwyth University. She is currently a postdoctoral researcher at the Royal Veterinary College where she is studying interactions within the gut microbiomes of chickens for their effects on performance phenotypes and public health. 
\\

\textbf{Anh Nguyen}
Anh Nguyen is a Senior Lecturer (Associate Professor) and directs the Smart Robotic Lab at the Department of Computer Science, University of Liverpool. He received my PhD from the Italian Institute of Technology (IIT), Italy. Previously, he worked at the Hamlyn Centre for Robotic Surgery, Imperial College London, Australian National University, Inria, and The University of Adelaide. He is an Associate Editor for IEEE Transactions on Medical Robotics and Bionics.
\\

\textbf{Dominic Richards}
Dom works in industrial applications of AI in the financial technology sector. He was previously AI Group Leader at the Hartree Centre. He holds a PhD and MSc from the University of Manchester, as well as an MEng and MSc from the University of Oxford.  
\\

\textbf{Dong Xia}
Dong Xia obtained his B.Sc. from Shandong University followed by M.Sc. by research in the University of Edinburgh and Ph.D. in the University of Liverpool on infection biology to characterise the proteomics of Toxoplasma gondii. He is currently Senior Lecturer in Bioinformatics at the Royal Veterinary College with research focuses on understanding the molecular mechanisms of host-pathogen interactions using multi-omics platforms coupled with bioinformatics and machine learning algorithms.
\\

\textbf{Ritesh Krishna}
Ritesh Krishna is an interdisciplinary technical leader at IBM Research with 20+ years of R\&D experience spanning across Omics technologies, Digital infrastructure, and Machine Learning.  Ritesh joined IBM Research in 2016 to kickstart the Computational Genomics programme with a focus on innovating around Data and AI challenges in multi-omics. Before that, Ritesh was associated with the Institute of Integrative Biology, Centre for Genomics Research and the Institute of Infection and Global Health at the University of Liverpool. Ritesh obtained his PhD in Computer Science from the University of Warwick and holds two master’s degrees from India and UK.

\vfill
\end{document}